\documentclass[12pt,onecolumn,draftclsnofoot]{IEEEtran}
\IEEEoverridecommandlockouts
\usepackage{amsmath,amsfonts,amsthm,amssymb,amsbsy,bm,paralist,color}
\usepackage{graphicx,algorithmic,algorithm}
\usepackage{cite}
\usepackage{lipsum}
\usepackage{cuted}
\usepackage{tikz}
\usetikzlibrary{decorations.pathreplacing}

\theoremstyle{definition}

\newtheorem{lem}{Lemma}[]

\theoremstyle{definition}

\theoremstyle{remark}

\newcommand\peqref[1]{(\mathcal{P}\ref{#1})}
\newcommand\fref[1]{Fig. \ref{#1}} 

\newcommand{\abs}[1]{\left\lvert#1\right\rvert}

\newcounter{prb}
\begin{document}
\title{Delay Minimization for Hybrid BAC-NOMA Offloading in MEC Networks}
\author{Haodong Li}
\maketitle
\begin{abstract}
This paper studies the offloading service improvement of multi-access edge computing (MEC) based on backscatter communication (BackCom) assisted non-orthogonal multiple access (BAC-NOMA). A hybrid BAC-NOMA protocol is proposed, where the uplink users backscatter their tasks by leveraging the downlink signal, and then the remaining data is transmitted through uplink NOMA. In particular, a resource allocation problem is formulated to minimize the offloading delay of uplink users. The non-convex problem is transformed into a convex problem, and an iterative algorithm is developed accordingly. Simulation results show that the proposed protocol outperforms the benchmark in terms of offloading delay.
\end{abstract}
\vspace{-0.5cm}
\section{Introduction}
%The explosive development of the Internet of Things (IoT) enables variant of novel applications with intensive computational capability\cite{9133107}.
 Multi-access edge computing (MEC) has been recognized as an essential technology in the envision of sixth-generation (6G) mobile network, where powerful servers are deployed at the edge of the based station (BS)\cite{9314080,LI2020241}. Hence, mobile devices are encouraged to offload their computational intensive tasks to the edge server and then obtain the processed results.
The MEC offloading demands significant communication service improvement to accommodate the large traffic volume, and recent works show that the combination of non-orthogonal multiple access (NOMA) and MEC can significantly improve the spectrum efficiency and reduce the latency\cite{9174768,9179779}. Moreover, backscatter communication (BackCom) can prolong the lifetime of energy constrained mobile devices, where other signals can be used to excite the circuit of the BackCom device (BD)\cite{9420716,9223730,9422161}. Therefore, the combination of BackCom and NOMA (BAC-NOMA) is considered to efficiently improve the offloading service of the resource limited devices.

In previous works, a few BAC-NOMA protocols have been applied to improve MEC offloading performance. For instance, the authors in \cite{9223730} proposed a BAC-NOMA scheme, where the NOMA users receive the signals from both the BS and the BD that excited by the downlink transmission. The full-duplex (FD) BAC-NOMA scheme was considered in \cite{9420716}, where the downlink transmission was used to excite the circuit of BDs to perform uplink transmission to the BS. The uplink sum rate maximization problem was studied and the optimal solution was obtained by reformulating the problem as a linear programming problem.

Different from the existing works which mainly focus on the EE maximization \cite{9223730} and sum rate maximization \cite{9420716,8851217} of BAC-NOMA schemes, this paper considers the delay minimization problem of a hybrid BAC-NOMA assisted MEC offloading scenario. In particular, the signal of the downlink transmission can excite the circuit of BDs to upload the tasks to the MEC server. Since the existence of FD interference, those BDs may not be able to finish offloading when the downlink transmission is active. Therefore, a second time duration is scheduled to continue the offloading by using the NOMA uplink transmission. By applying the Dinkelbach method and quadratic transformation, an iterative based algorithm is proposed to obtain the sub-optimal solution to the original non-convex problem. By comparing with benchmarks, simulation results demonstrate that the proposed protocol can significantly reduce the offloading latency.
\section{System Model and Problem Formulation}
\subsection{System Model}
Consider a BS equipped with an MEC server to serve one downlink user $U_0$ and $k$ uplink BDs which is denoted by $\mathrm{BD}_k$, where $k \in \{1,\ldots,K\}$. Assume that during the time slot $t_0$, the BS starts the transmission to $U_0$. The downlink transmit power is denoted by $P_0$, and the received signal at $\mathrm{BD}_k$ can be expressed as $\sqrt{P_0}h_k s_0$, where $h_k$ is the channel gain between the BS and $\mathrm{BD}_k$, and $s_0$ represents the downlink signal to $U_0$ with unit power. By leveraging the BackCom protocol, the downlink signal can excite the circuits of those BDs, and the reflected signal from $\mathrm{BD}_k$ can be expressed as $\sqrt{\eta_k P_0} h_k s_0 s_k$ where $\eta_k \in [0,1]$ is the BackCom reflection coefficient, and $s_k$ denotes the signal of $\mathrm{BD}_k$ with unit power. 
Therefore, the received reflected signal at the BS can be represented by \cite{9420716}:
\begin{equation}
	y_{BS} = \sum_{k=1}^{K} \sqrt{\eta_k P_0} h^{2}_k s_0 s_k + I_{SI} + n_{BS},
\end{equation}
where $I_{SI}$ represents the FD self-interference, which is assumed to be complex Gaussian distributed, and it satisfies $I_{SI} \sim \mathcal{CN}(0,\alpha P_0 \abs{h_{SI}}^2)$, and $\alpha \in [0,1]$ represents the amount of residual interference. The complex additive white Gaussian noise (AWGN) at the receiver is denoted by $n_{BS} \sim \mathcal{CN}(0,\sigma^2)$ with zero mean and variance $\sigma^2$. 

Without loss of generality, the BDs are arranged in a descending order of the effective channels. The successive interference cancellation (SIC) technique is adopted for NOMA decoding, in which the BS decodes the signals of BDs with better channel gains priorly and treats the signals of the BDs with worse channel gains as interference. By assuming the downlink signal $s_0$ is perfectly known by the BS, the achievable rate of $\mathrm{BD}_k$ can be expressed as 
\begin{equation} \label{r_b}
\begin{split}
	&r_{b,k} = \\&B\log_2 \left(1+ \frac{\eta_k P_0 \abs{h_k}^4 \abs{s_0}^2}{\sum_{j=k+1}^{K} \eta_j P_0 \abs{h_j}^4\abs{s_0}^2 +\alpha P_0 \abs{h_{SI}}^2+\sigma^2} \right),\raisetag{45pt}
\end{split}
\end{equation}
where $0\leq\alpha\ll 1$ denotes the amount of FD residual self-interference \cite{9420716}. 
Moreover, since the characteristics of BackCom transmission, the reflected signals from the BDs will introduce extra interference to the downlink user, and the received signal at $U_0$ can be written as\cite{9420716}:
\begin{equation}
	y_{0} = \sqrt{P_0}h_0 s_0 + \sum\nolimits_{k=1}^{K}\sqrt{\eta_k P_0}g_k h_k s_0 s_k + n_{U_0},
\end{equation}
where $h_0$ is the channel gain between the BS and $U_0$, $g_k$ represents the channel gain between $U_0$ and $\mathrm{BD}_k$, and $n_{U_0}\sim\mathcal{CN}(0,\sigma^2)$ denotes the AWGN between $U_0$ and the BS.
Hence, the achievable rate for $U_0$ is
\vspace{-0.3cm}
\begin{equation}
	r_0 = B\log_2 \left(1+ \frac{P_0 \abs{h_0}^2}{\sum_{k=1}^{K} \eta_k P_0 \abs{g_k}^2 \abs{h_k}^2 +\sigma^2} \right).
\end{equation}
\vspace{-0.3cm}
\par
On top of the BackCom transmission, this paper also considers the active uplink transmission scheme, in which an extra time duration can be designated for BDs' uplink transmission. It is assumed that the offloading may not be able to finish within $t_0$. Forcing the BDs to finish offloading within $t_0$ could require a higher downlink power $P_0$ or a longer transmission time $t_0$, which may degrade the transmission service of $U_0$. Instead, an additional time duration $t_a$ can be utilized by the BDs to perform NOMA uplink transmission.
The achievable active transmission rate for $\mathrm{BD}_k$ can be expressed as
\begin{equation}
	r_{a,k} = B\log_2 \left(1+\frac{p_{a,k} \abs{h_k}^2}{\sum_{j=k+1}^{K} p_{a,j}\abs{h_j}^2+ \sigma^2} \right),
\end{equation}
where $p_{a,k}$ denotes the active transmission power of $\mathrm{BD}_k$.
\subsection{Problem Formulation}
The aim of this paper is to minimize the MEC offloading delay of $K$ BDs. The offloading delay is defined as $T=t_0+t_a$, where $t_0$ is assumed to be fixed and predetermined by the BS since it is primarily assigned to the downlink transmission for $U_0$. Consider $K$ BDs offload tasks to the MEC server through NOMA simultaneously, the time consumption $t_a$ for offloading the remaining data can be calculated as
	\begin{equation}
		t_a=\frac{L_k-t_0 r_{b,k}}{r_{a,k}}, k \in \{1,...,K\}.
	\end{equation}
	where $L_k$ is the offloading data amount of $\mathrm{BD}_k$.
%%%%%%%%%%%%%%%%%%%%%%%%%%%%%%%%%%%%%
Hence, based on the expression of $t_a$, the delay minimization problem can be formulated as
\begin{subequations}\refstepcounter{prb}
	\label{P1}
	\begin{align}
		\peqref{P1} &\min_{P_0, \bm{p}_a , \bm{\eta}}		&&  t_0 + t_a	\\
		&\mathrm{s.t.}	&& r_0 \geq \gamma_0	,	\label{C1}	\\
		&&&0\leq t_{a} p_{a,k}  \leq E_{k,\max}, \ \forall k	 \label{C2}\\
		&&& 0\leq p_{a,k} \leq P_{a,\max}, \forall k \label{C3}\\
		&&&  0 \leq P_0 \leq P_{0,\max},\label{C4}\\
		&&& 0 \leq \eta_k \leq 1,\quad \forall k \label{C5}
	\end{align}
\end{subequations}
where $\bm{p}_a \triangleq [p_{a,1}, p_{a,2},...,p_{a,k}]$, and $\bm{\eta} \triangleq [\eta_{1},\eta_{2},...,\eta_{K}]$. 
The minimum achievable rate for the downlink transmission is guaranteed by \eqref{C1}. Constraint \eqref{C2} ensures that the the energy consumption for offloading is limited by the amount of energy of each BD. The power of uplink transmission for $\mathrm{BD}_k$ is limited in \eqref{C3}. Constraint \eqref{C4} guarantees the range of the downlink transmission power, and \eqref{C5} limits the range of the reflection coefficient.  Since $\peqref{P1}$ is non-convex due to the fractional form of the objective function and coupled variables, a convex equivalent of this problem is developed to tackle the resource allocation problem in the next section.

\section{Problem Transformation and Solution}
In this section, the paper proposes two approaches to transform the original non-convex problem into two equivalent convex forms in order to adapt the two different offloading scenarios, i.e., pure BAC-NOMA offloading and hybrid BAC-NOMA offloading. If BDs are able to finish offloading within $t_0$, the pure BAC-NOMA scheme becomes feasible, and the second time duration $t_a = 0$. Otherwise, the hybrid BAC-NOMA scheme is adopted.
 Based on the above two schemes, an iterative based algorithm is proposed to obtain the resource allocation efficiently.
\begin{lem} \label{lem1}
	Define $\tilde{L}=\sum^{K}_{k=1} L_k$, $t_a$ is equivalent to the following expression:
\begin{equation}
		t_a = \frac{\tilde{L} -t_0 R_b}{R_a},
	\end{equation}
	where $R^{b} =  \sum^{K}_{k=1} r_{b,k}$ and $R^{a} =  \sum^{K}_{k=1} r_{a,k}$.
\end{lem}
\begin{proof}
	First, the remaining data after the first time duration $t_0$ is denoted as $\beta_{a,k} = L_k - t_0 r_{b,k}$. As all BDs finish offloading their tasks simultaneously within the same transmission time $t_a$, the following relation holds:
	\begin{equation}
		t_a=\frac{\beta_{a,1}}{r_{a,1}}=\frac{\beta_{a,2}}{r_{a,2}}=...=\frac{\beta_{s,K}}{r_{a,K}}
	\end{equation}
	Since $\frac{\beta_{a,1}}{r_{a,1}} = \frac{\beta_{a,k}}{r_{a,k}}$, it can be found that $\epsilon_k \triangleq \frac{\beta_{a,k}}{\beta_{a,1}}=\frac{r_{a,k}}{r_{a,1}}$. Then, the following fraction can be constructed by multiplying the same term with both numerator and the denominator:
	\begin{equation}
		\frac{\beta_{a,1}}{r_{a,1}}=\frac{\beta_{a,1}\left(1+ \epsilon_2 +\epsilon_3 +...+ \epsilon_K \right)}{r_{a,1}\left(1+ \epsilon_2 +\epsilon_3 +...+ \epsilon_K \right)}.
	\end{equation}
	Therefore, the above equation can be written as
	\begin{equation} \label{lem1_eq1}
		\begin{split}
			\frac{\beta_{a,1}}{r_{a,1}}=&\frac{\beta_{a,1}\left(1+ \frac{\beta_{a,2}}{\beta_{a,1}} +...+ \frac{\beta_{a,K}}{\beta_{a,1}} \right)}{r_{a,1}\left(1+ \frac{r_{a,2}}{r_{a,1}} +...+ \frac{r_{a,K}}{r_{a,1}} \right)}=\frac{\sum^{K}_{k=1}\beta_{a,k}}{\sum^{K}_{k=1}r_{a,K}},
		\end{split}
	\end{equation}
	and the transmission time $t_a$ can be written as
	\begin{equation}\label{ta_org}
		\begin{split}
			t_a =& \frac{\sum^{K}_{k=1} \left( L_k -t_0 r_{b,k}\right)}{\sum^{K}_{k=1}r_{a,k}}\\
			%=&\frac{\sum^{K}_{k=1} L_k -t_0 B\log_2\left( 1 + \frac{\sum^{K}_{k=1}\eta_{k}P_{0} \abs{h_{k}}^4\abs{s_0}^2}{\alpha P_{0} \abs{h_{SI}}^2 + \sigma^2}\right)}{B\log_2 \left(1 + \frac{\sum_{k=1}^{K}p_{a,k}\abs{h_{k}}^2}{\sigma^2} \right)}.\qedhere
			%=&\frac{\tilde{L} -t_0 R_b}{R_a}.
		\end{split}\raisetag{20pt}
	\end{equation}
	Based on \eqref{r_b}, the sum rate of $\mathrm{BD}_K$ and $\mathrm{BD}_{K-1}$ is
\begin{equation} \label{sum_01}
	\begin{split}
		 r_{b,K} &+ r_{b,K-1} =\\
		 &B \log_2 \left(1+ \frac{\eta_{K-1} P_0 \abs{h_{K-1}}^4 \abs{s_0}^2 + \eta_{K} P_0 \abs{h_{K}}^4 \abs{s_0}^2}{\alpha P_0 \abs{h_{SI}}^2+\sigma^2}  \right).\raisetag{45pt}
	\end{split}
\end{equation}
Similarly, it can be found that the sum rate of 
\begin{equation}
	\begin{split}
		r_{b,K-2} +& r_{b,K-1} + r_{b,K} =\\ &B \log_2 \left(1+ \frac{\sum_{i=K-2}^{K} \eta_{i} P_0 \abs{h_{i}}^4 \abs{s_0}^2}{\alpha P_0 \abs{h_{SI}}^2+\sigma^2}  \right).
	\end{split}
\end{equation}
Therefore, by repeating the above manipulations, the sum rate of $K$ BDs using the backscatter transmission can be written as
\begin{equation} \label{sumrate_bc}
	R^{b} =  \sum^{K}_{k=1} r_{b,k} = B\log_2\left( 1 + \frac{\sum^{K}_{k=1}\eta_{k}P_{0} \abs{h_{k}}^4\abs{s_0}^2}{\alpha P_{0} \abs{h_{SI}}^2 + \sigma^2}\right). 
\end{equation}
Similar to \eqref{sumrate_bc}, the sum rate for the active uplink transmission can be written as
\begin{equation} \label{sumrate_at}
	R^{a} =\sum^{K}_{k=1} r_{a,k} = B\log_2 \left(1 + \frac{\sum_{k=1}^{K}p_{a,k}\abs{h_{k}}^2}{\sigma^2} \right).
\end{equation}
By substituting \eqref{sumrate_bc} and \eqref{sumrate_at} into \eqref{ta_org}, the offloading latency $t_a$ can be written as $t_a = \frac{\tilde{L} -t_0 R_b}{R_a}$.
\end{proof}
\subsection{Pure BAC-NOMA Case}
If all BDs can finish offloading within $t_0$, the second time duration $t_a$ will be zero. By assuming the back-scatter transmission satisfies the constraint $L_{k}=t_0 r_{b,k}$.
According to Lemma \ref{lem1}, the above constraint is equivalent to $\tilde{L}=t_0 R_b$.
Hence, the original problem becomes a feasibility problem, which can be formulated as
\begin{subequations}\refstepcounter{prb}
	\label{P1p}
	\begin{align}
		&\peqref{P1p}\ &&\mathrm{find}\quad P_0 , p_{r,k}	\nonumber \\
		&\mathrm{s.t.}	&& \hspace{-0.6cm} \sum^{K}_{k=1}p_{r,k} \abs{h_{k}}^4\abs{s_0}^2- \gamma\left(\alpha P_{0} \abs{h_{SI}}^2 + \sigma^2\right) = 0,\label{P1p-C1}\\
		&&&\hspace{-0.6cm} P_0\abs{h_0}^2 - \tilde{\gamma_0} \sum\nolimits^{K}_{k=1}p_{r,k} \abs{g_k}^2 \abs{h_k}^2 -\tilde{\gamma_0}\sigma^2 \geq 0,  \label{P1p-C2}\\
		&&&\hspace{-0.6cm}  0 \leq P_0 \leq P_{0,\max},\label{P1p-C3}\\
		&&& \hspace{-0.6cm} 0 \leq p_{r,k} \leq P_0,\quad \forall k, \label{P1p-C4}
	\end{align}
\end{subequations}
where $p_{r,k} = \eta_k P_0$, $\gamma =  2^{\frac{\tilde{L}}{t_0 B}-1}$, and $\tilde{\gamma_0} = 2^{\frac{\gamma_0}{B}}-1 $.
It is evident that the above problem is convex and the feasibility can be found by some optimization solvers, such as CVX.
\subsection{Hybrid BAC-NOMA Case}
 When $ \tilde{L} \geq t_0 R_b$, which means BDs cannot finish offloading during $t_0$, and the remaining data will be offloaded during $t_a$. Since the objective function in $\peqref{P1}$ is fractional and non-convex, the Dinkelbach method \cite{dink1967} is leveraged to transform the original problem into an equivalent objective function. 
The principle of the Dinkelbach's method is to convert the fractional format into the following optimization problem:
\begin{subequations}\refstepcounter{prb}
	\label{P2}
	\begin{align}
		\peqref{P2} &\min_{P_0, \bm{p}_a , \bm{\eta}}		&& 
		\begin{aligned}
		\tilde{L} &-t_0 R_b -\mu R_a
		\end{aligned} \label{P2:obj}  \\
		&\mathrm{s.t.}	&& 0\leq \mu p_{a,k} \leq E_{k,\max} \label{P2:C1} \\
		&&&\eqref{C1} ,\eqref{C3}, \eqref{C4}, \eqref{P1p-C4} \nonumber
	\end{align}
\end{subequations}
where $\mu$ is an auxiliary variable which is determined by the proposed iterative algorithm in the later section. 
%Refer to the principle of the Dinkelbach method \cite{dink1967}, the optimal EE $\mu^*$ can be achieved if the following equation can be satisfied:
%\begin{equation}
%	\min_{k \in \mathcal{K}} \quad r^{b}_{k}(\eta^*_k,P^*_0)+r^{a}_{k}({p_{a,k}}^*) + r^{l}_{k}({f^{l}_{k}}^*)-\mu^*\left({p_{a,k}}^*+{p_{l,k}}^*\right) = 0,
%\end{equation}
%where a superscript $*$ denotes that optimization variable is the optimal solution to the problem. The approach to find those optimal solutions can be developed by an iterative algorithm based on the Dinkelbach method.

Problem $\peqref{P2}$ is still non-convex due to the existence of FD interference in the second term in $R_b$. Hence, to further simplify the problem, quadratic transform \cite{8314727} is applied to the objective function of $\peqref{P2}$, and the reformulated problem can be expressed as
\begin{subequations}\refstepcounter{prb}
	\label{P3}
	\begin{align}
		&\min_{\substack{P_0, \bm{p}_a, \bm{p}_{r}}}		
		\hspace{-1cm}&&  \begin{aligned}
			\hspace{-0.5cm}\peqref{P3} \quad F&(\mu) = \tilde{L}-\mu B\log_2 \left(1 + \frac{\sum_{k=1}^{K}p_{a,k}\abs{h_{k}}^2}{\sigma^2} \right) \\
			&- t_{0}B\log_2 \left\{\begin{aligned}
				1 + 2y&\sqrt{\sum^{K}_{k=1} p_{r,k} \abs{h_{k}}^4\abs{s_0}^2} \\&-y^2\left( \alpha P_{0} \abs{h_{SI}}^2 + \sigma^2\right)
			\end{aligned}\right\}
		\end{aligned}	\\
		&\mathrm{s.t.}	&&  \eqref{C3},\eqref{C4},\eqref{P1p-C2}, \eqref{P1p-C4},\eqref{P2:C1}\nonumber
	\end{align}
\end{subequations}
where $y$ is an auxiliary variable which will be updated iteratively.
For given $p_{r,k}$ and $P_0$, based on the principle of quadratic function, the optimal $y^*$ to minimize the objective function can be determined by 
\begin{equation} \label{eq-y}
	y^* = \frac{\sqrt{\sum^{K}_{k=1} p_{r,k} \abs{h_{k}}^4\abs{s_0}^2}}{\alpha P_{0} \abs{h_{SI}}^2 + \sigma^2}.
\end{equation} 

Hence, for fixed auxiliary variables $y$ and $\mu$, problem $\peqref{P3}$ becomes a convex problem, and CVX can be utilized to solve this problem efficiently. By providing an initial feasible $y^{(0)}$, the optimal $P^{*}_0$, $p^{*}_{r,k}$, and $p^{*}_{a,k}$ can be obtained based on the given $y^{(0)}$. Then, $y$ can be updated for next iteration based on \eqref{eq-y} with the obtained optimization variables.

Therefore, an algorithm is proposed to iteratively update the power allocation solution and the iterative variable $\mu$. The procedure of this proposed scheme is summarized in Algorithm \ref{algorithm1}. With the increment of $l$, the Dinkelbach iterative variable $\mu$ will finally converge to the $\varepsilon$-optimal $\mu^*$, where the $\varepsilon$-optimal $\mu^*$ can be achieved if $F(\mu^*)	\leq \varepsilon$. The convergence of Dinkelbach based algorithm was proved in \cite{dink1967}, and simulation results also show that the proposed algorithm is able to converge within a few iterations, which is presented later in the next section.
\begin{algorithm}
	\caption{Iterative based resource allocation algorithm}
	\label{algorithm1}
	\begin{algorithmic}[1]
		%\STATE Input: $h_k$, $g_k$, $L_k$, $E_{k,\max}$, $\gamma_0$, $t_0$, $\varepsilon$
    	\STATE Set $l=0$, $\mu = +\infty$    	
    	\IF{$\peqref{P1p}$ is feasible}
    	\STATE Pure backscatter transmission is adopted.
    	\STATE Solve $\peqref{P1p}$ and obtain $P_0^*$ and $p_{r,k} ^*$.
		\ELSE
    	\STATE Initialize $P_0$, $p_{r,k}$ with feasible values.
    	\REPEAT
    	\STATE $l = l+1$
    	\STATE Update $y^{(l)}$ by \eqref{eq-y}.
    	\STATE Solve problem $\peqref{P3}$ with fixed $\mu$, and obtain $P_0^{(l)}$, $p_{r,k}^{(l)}$, and $p_{a,k}^{(l)}$.
    	\STATE Solve $F(\mu^{(l)})$.
    	\vspace{-0.4cm}
    	\STATE  updata $\mu$ as $ \mu^{(l)}=\frac{\tilde{L} -t_0 B\log_2\left( 1 + \frac{\sum^{K}_{k=1}p_{r,k}^{(l)} \abs{h_{k}}^4\abs{s_0}^2}{\alpha P_{0}^{(l)} \abs{h_{SI}}^2 + \sigma^2}\right)}{B\log_2 \left(1 + \frac{\sum_{k=1}^{K}p_{a,k}^{(l)}\abs{h_{k}}^2}{\sigma^2} \right)}$.
    	    	\vspace{-0.4cm}
    	\UNTIL{{$F(\mu^{(l)}) \leq \varepsilon$}.}
    	\STATE  $P_0^* = P_0^{(l)}$, $p_{r,k}^{*}=p_{r,k}^{(l)}$, and $p_{a,k}^{*}=p_{a,k}^{(l)}$.
    	\ENDIF
	\end{algorithmic}
\end{algorithm}
\section{Simulation Results}
In this section, simulation results are presented and discussed to evaluate the proposed hybrid NOMA-BackCom scheme. 
The simulation settings are listed as follows. The channel gain from the BS to $\mathrm{BD}_k$ is $h_{k} = \tilde{h}_{k} d^{-\frac{\chi}{2}}$, where $\tilde{h}_{k}$ is a Rayleigh fading channel coefficient and $d_k$ is the corresponding distance between  $\mathrm{BD}_k$ and the BS, and $\chi$ denotes the path-loss exponent which is set to $3.76$ throughout the simulation. Similarly, the channel gain $h_0$ that between the BS and $U_0$, and the channel gain $g_k$ that between $U_0$ and $\mathrm{BD}_k$ are realized in the same way, in which the distance between $U_0$ and $\mathrm{BD}_k$ is denoted by $\tilde{d}_k$. The transmission radius of the BS is $50$ m, and all $\mathrm{BD}$s as well as $U_0$ are covered within the disc region. The AWGN power is $\sigma^2 = BN_0$, where $B=5$ MHz, and $N_0 = -94$ dBm/Hz denotes the AWGN spectrum density. The error tolerance $\varepsilon = 10^{-4}$, and the simulation results are obtained after $10^{3}$ realizations.

\begin{figure}[!t]	
	\centering
	\includegraphics[scale=0.6,trim=7cm 9cm 7cm 9cm]{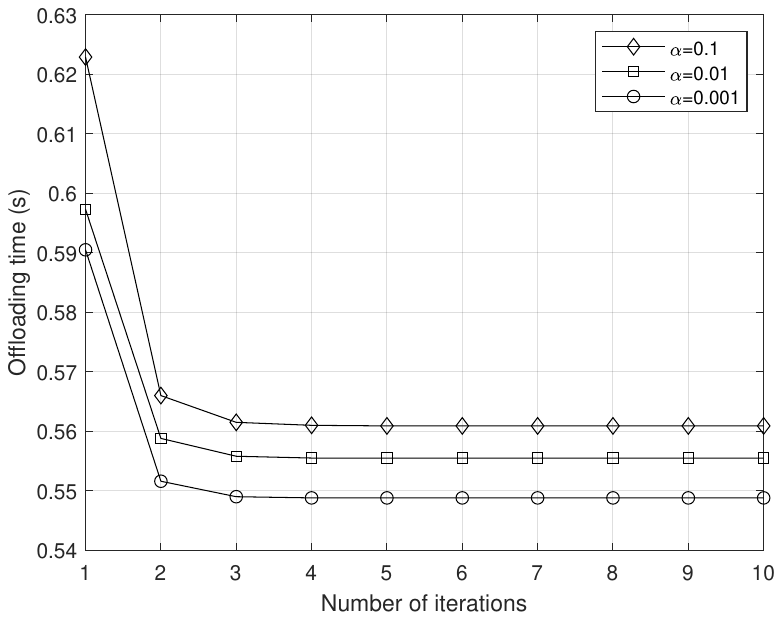}
	\caption{The convergence performance and delay comparison.}\label{r_1}
\end{figure}

The convergence performance of the proposed Algorithm \ref{algorithm1} and the effect of amount of FD residual self-interference are demonstrated in \fref{r_1}. In this figure, the data length of each user $L_k$ is set to $10^6$ bits, $t_0 = 0.5$ s, and the maximum uplink transmission power $p_{a,\max}=0.5$ watt. According to the simulation results, the algorithm can efficiently converge within a few iterations. Additionally, the amount of FD residual self-interference can affect the offloading delay since the higher residual can reduce the achievable rate of each BD and hence the remaining data amount for offloading in the second duration increases.
\begin{figure}[!t]	
	\centering
	\includegraphics[scale=0.6,trim=7cm 9cm 7cm 9cm]{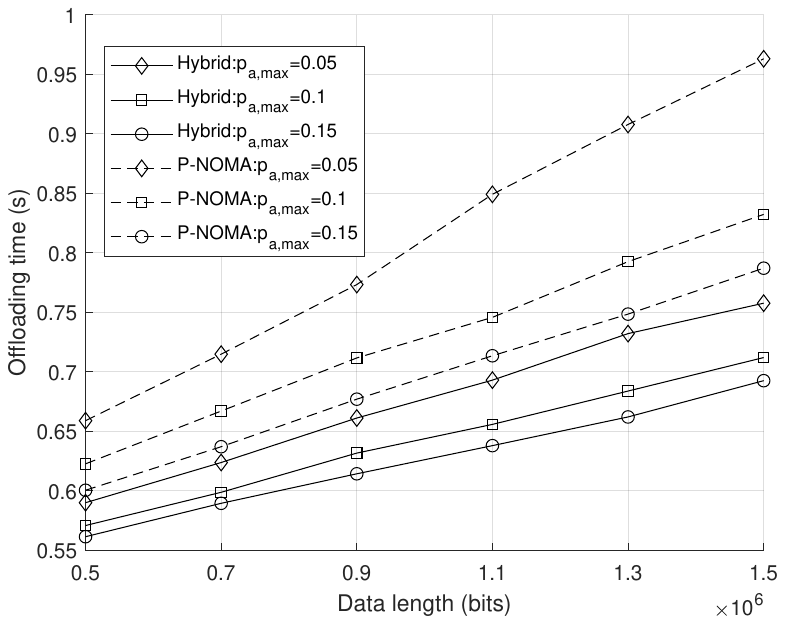}
	\caption{Average offloading delay by hybrid offloading and pure NOMA uplink transmission as a function of offloading data length.}\label{r_2}
\end{figure}

%\begin{figure}[!t]	
%	\centering
%	\includegraphics[scale=0.5,trim=7cm 9cm 7cm 9.3cm]{r_3.pdf}
%	\caption{Average offloading energy consumption (a) and average offloading delay (b) as functions of offloading data length.}\label{r_3}
%\end{figure}
As shown in \fref{r_2}, the relationship between the average offloading delay and the data length of each user is studied under the fixed QoS constraint $\gamma_0=2$ Mbit$/$s and $E_{\max}=0.1$ joules. The pure NOMA scheme is considered as a benchmark, where BDs do not perform BackCom during $t_0$, which means that $\eta_k = 0,\ \forall k$.
For both transmission schemes, the average delay increases with the growth of offloading data length. The dashed lines which represent the pure NOMA uplink transmission show higher average offloading delay compared to the solid line with the hybrid offloading scheme. The proposed hybrid scheme is efficient for offloading delay reduction. Moreover, the higher power budget for offloading can significantly reduce the system delay for both cases.

%In \fref{r_3}, the maximum energy budget for each user $E_{k,max}= 0.1$ joules. \fref{r_3}(a) depicts the relationship between the offloading data length and the energy consumption. As more data being offloaded to the BS, each device consumes more energy for uplink transmission, until the energy is used up for each device, and the task offloading becomes infeasible due to the constrained energy if more than that amount of data need to be uploaded. Comparing the curve with $p_{a,\max}=0.05$w, the curves with higher uplink transmit power budget $p_{a,\max}$ consume more energy for offloading the same amount of data. The reason is that, since the objective function is to minimize the offloading delay, each user tends to utilize higher transmit power as shown in \fref{r_3}(b). The energy consumption is monotonically increasing with the  transmit power $p_{a,k}$, which verifies the proposition.
\section{Conclusion}
In this work, the delay minimization problem for hybrid BAC-NOMA assisted MEC network is investigated. The original delay minimization problem is non-convex. By utilizing the Dinkelbach method and quadratic transformation, the non-convex problem is equivalently reformulated into a convex form. Simulation results demonstrated the performance superiority of the proposed hybrid scheme by comparing with the pure uplink NOMA scheme in terms of average delay.
\bibliographystyle{IEEEtran}
\bibliography{mybib}

\end{document}